\documentclass[prl,a4paper,onecolumn,preprint,footinbib,showpacs]{revtex4}
\usepackage{amsmath,bm}

\newcommand{\qo}[1]{\textquotedblleft #1\textquotedblright}

\begin{document}
\title{The  geometrical nature of  the cosmological  inflation in the framework  of  the  Weyl-Dirac conformal gravity theory}
\author{Francesco De Martini}
\affiliation{Accademia dei Lincei, via della Lungara 10, I-00165 Roma, Italy}
\email{francesco.demartini@uniroma1.it}
\author{Enrico Santamato}
\affiliation{Dipartimento di Fisica, Universit\`{a} Federico II, Napoli 80126}
\email{enrico.santamato@na.infn.it}
\date{version: February 2  2017}
\begin{abstract}
The nature of the scalar field responsible for the cosmological inflation, the \qo{inflaton}, is found to be rooted in the most fundamental concept of the Weyl's differential geometry: the parallel displacement of vectors in curved space-time. The Euler-Lagrange theory based on a scalar-tensor Weyl-Dirac Lagrangian leads straightforwardly to the Einstein equation admitting as a source the characteristic energy-momentum tensor of the inflaton field. Within the dynamics of the inflation, e.g. in the slow roll transition from a \qo{false} toward a \qo{true vacuum}, the inflaton's geometry implies a temperature driven symmetry change between a highly symmetrical \qo{Weylan} to a low symmetry \qo{Riemannian} scenario. Since the dynamics of the Weyl curvature scalar, constructed over differentials of the inflaton field, has been found to account for the quantum phenomenology at the microscopic scale, the present work suggests interesting connections between the \qo{micro} and the \qo{macro} aspects of our Universe.
\end{abstract}
\maketitle
A huge step forward in theoretical cosmology, and today a very important landmark of modern science, was the proposal by A. Starobinsky followed one year later by A.Guth, A. D. Linde, A. Albrecht and P.J. Steinhard~\cite{Starobinsky1980, Guth1981,Linde1982,AlbrechtSteinhardt1982} of the \textit{inflation}, an epoch of fast accelerating expansion of the early Universe that caused the Universe to expand through about 70 $e$-folds in a very small fraction of a second. This expansion, driven by a scalar field called \qo{inflaton}, was originally argued to solve the problem of why the universe is so smooth at large scales. Moreover it later turned out to provide a consistent solution also to a host of different crucial problems among which the growth of structures in the Universe arising from magnified quantum fluctuations, the no observation of magnetic monopoles, the isotropy of the cosmic microwave background etc.~\cite{Dodelson2003,Weinberg1972}.  Over the years, the undeniable success of this idea was however somewhat tarnished and even questioned by the failure of finding the physical mechanism underlying the fundamental nature of the \qo{inflaton} concept. In the present letter it is claimed that the fundamental nature of this scalar field is indeed \textit{geometrical}, based on the conformal differential geometry introduced by Hermann Weyl two years after the publication of the first Einstein paper on General Relativity (GR)~\cite{Weyl1952,Lord1979}. This geometry rests on the following general statement: \qo{\textit{All Laws of Physics are invariant under} (A) \textit{any change of coordinates including time and} (B) \textit{under any change of calibration}}. The first part, (A) expresses the well known \textit{covariance} of the GR based on the Riemann geometry.  The second part (B), a significant addition by Weyl to Riemann's, expresses the \textit{conformal} property of the metric space-time theory. Accordingly, the above statement is usually cast in a simpler form: \qo{\textit{All Laws of Physics are conformally-covariant}} (or \textit{Weyl covariant} or, in short, \textit{co-covariant}).\\

The rationale of the Weyl geometry can be outlined as follows. Taking Planck's constant ($\hbar$) and the velocity of light ($c$) to be constant by definition, any physical quantity $X$, e.g. represented by a scalar, a tensor, a spinor etc, can be assigned a unit that is a power of a physical length $\tilde L$: $X\rightarrow\tilde L^{W(X)}$.  Furthermore, to that quantity can be assigned a transformation law $X\rightarrow e^{W(X)\lambda(x)}X$ under a conformal mapping. $W(X)$ is  a (positive or negative) real number dubbed \qo{\textit{Weyl weight of} $X$} (or \qo{dimensional number of $X$}) and $\lambda(x)$  is a regular, real  function of the space-time coordinates. Thus, the conformal mapping is a \qo{unit transformation} amounting to a local space-time redefinition of the unity of length, i.e. of the \qo{\textit{calibration}}. This concept is rooted into the most basic operation of the differential geometry i.e. the \qo{parallel displacement} of any vector in a non-Euclidean manifold.  In fact, according to Weyl, the parallel displacement from two infinitely nearby points $P$ and $P + dP$ in spacetime acts on the length $\ell$ of any vector by inducing a calibration change  $\delta\ell=\ell\phi_\rho dx^\rho$, where $\phi_\rho$ is a universal \qo{\textit{Weyl vector}}, defined in the whole space-time. In summary, the metric structure of the Weyl geometry implies two fundamental forms: the \textit{quadratic} Riemannian one, $g_{\rho\sigma}dx^\rho dx^\sigma$, being $g_{\rho\sigma}=g_{\sigma\rho}$ the metric tensor, and the Weyl \textit{linear} one $\phi_\rho dx^\rho$. In Riemann's geometry is always $\phi_\rho=0$. The parallel displacement is \textit{integrable} iff a scalar \qo{\textit{Weyl potential}} $\phi$ exists such as $\phi_\rho=\partial_\rho\phi$. As we shall see, the \qo{\textit{Weyl vector}} $\phi_\rho(x)$ and the corresponding scalar \qo{Weyl potential} $\phi(x)$, both defined in the whole space-time spanned by the $x^\rho$ coordinates play a basic role in the present work: indeed the \qo{inflaton} is identified with $\phi$~\cite{foot:Weyl}.\\

We first note that under a physics perspective the Weyl geometry indeed consists of an abelian, local, scale-invariance gauge theory implying the following group of transformations
\begin{eqnarray}\label{eq:gauge}
   \phi_\rho &\rightarrow& \phi_\rho + \partial_\rho\lambda(x) \nonumber \\
   g_{\rho\sigma}&\rightarrow& e^{2\lambda}g_{\rho\sigma}.
\end{eqnarray}
The "inflaton" $\phi$  is a "gauge field"~\cite{Weyl1952,Quigg1983}. The insightful perspective offered by this theory was supported by P.A.M. Dirac in a 1973 seminal paper~\cite{Dirac1973,foot:Weylerror}:\\
\qo{\textit{There is a strong reason in support of the Weyl's theory. It appears as one of the fundamental principles of Nature that the equations expressing basic laws should be invariant under the widest possible group of transformations. The confidence that one feels in Einstein GR theory arises because its equations are invariant under the wide group of transformations of curvilinear coordinates in Riemann space. The passage to Weyl geometry is a further step in the direction of widening the group of transformations underlying the physical laws. One has to consider transformations [...] which impose stringent conditions on them}}.\\
The \qo{stringent conditions} alluded by Dirac imply in the first place the correct choice by definition of several constant units in terms of which the physical quantities are measured: these units must be mutually independent in the sense that a dimensionless number cannot be constructed with them. As in the relativistic quantum theory, it is conventional to take $\hbar,c,m_e$ (the electron mass) to be \textit{constant by definition}~\cite{Lord1979,Quigg1983}. Other gauges, e.g. by replacing $m_e$ by the \qo{gravitational constant} $G$, lead in general to different theories which are mutually connected by conformal mapping. Furthermore, the Lagrangian density $L$  from which the dynamical equations are derived as well any measurable quantity $X$ attained as result of the theory must be co-covariant, and then expressed by the zero value of the Weyl weight $W(L)=0$, $W(X)=0$. At the end of the calculation a careful application of the initial and boundary conditions must be undertaken in order to select the physically admissible results among the special solutions of the theory. The special solutions have the property that particle masses are independent of positions in space-time and the gravitational constant $G$ is a true constant. Thus the statement that all electrons have the same mass, all hydrogen atoms have the same size etc. may be taken to be true by definition. All this circumvents the original criticism to the Weyl theory~\cite{Weyl1952,J.EhlersA.Schild1972,EhlersSchild1973,Trautman2012,DeLaurentisFatibeneFrancaviglia2014,Scholz2011,Scholz2012,FatibeneGarrutoPolistina2015}.  As it will be  shown below, the requirement of full covariance under conformal mapping is necessary only to determine the formal structure of the theory.\\

The implied meaning of the present work may be further enlightened by the fact that after 1970 several axiomatic approaches for deducing the projective and conformal structure of space-time  were carried out by using basic concepts such as light rays  and freely falling particles~\cite{J.EhlersA.Schild1972,EhlersSchild1973,Trautman2012,DeLaurentisFatibeneFrancaviglia2014,Scholz2011,Scholz2012,FatibeneGarrutoPolistina2015}. The higly remarkable fact is that all these investigations ended by assigning a Weylan, not a Riemannian structure to space-time. This raised a long lasting, and as yet unresolved interesting dilemma that could be understood by the possibility of the onset of a cosmological "phase transition" by which the now perceived Riemannian structure is induced via a dynamical process that breaks the Weyl invariance of the vacuum at some critical temperature $T_c$. This symmetry changing process is now included in the modern inflation theory, as it will be immediately shown.\\

All these concepts will be applied to a very general Weyl-Dirac conformal scalar-tensor theory involving the mass $m_E$ of an elementary particle. In order to keep the conformal structure of the theory, the mass of the particle  is expressed in form of a "mass field" $m_E\rightarrow{k_E\mu(x)}$, being the dimensionless coupling constant: $k_E$ an intrinsic particle's property, and $\mu(x)$  a scalar field, function of space-time, with weight $W(\mu)=-1$ ~\cite{Jammer2000}. The gravitational �constant� $G$ appearing in the Einstein gravitational equation has $W(G)=+2$ and so it cannot be regarded as constant in the  present approach. We define a dimensionless constant: $\alpha$ and assume that the ratio $l_C/l_P$ between the particle�s Compton length $l_C=\hbar/mc$ and the Planck length is independent of position in space and time. By applying the Dirac's Lagrange-multiplier method, the simplest form of the co-covariant Lagrangian density in $D=4$ can be expressed in the scalar-tensor form~\cite{Dirac1973,Lord1979}:
\begin{equation}\label{eq:L}
L = \sqrt{-g}\left\{\alpha\mu^2\left[\bar R +2 V(T,\phi,\mu)\right]-{D_{\rho}\mu}{D^{\rho}\mu}-\frac{\beta^{2}}{4}\phi_{\rho\sigma}\phi^{\rho\sigma}\right\}
\end{equation}
where $\bar R=R_g+R_W$ is the overall Riemann-Weyl curvature scalar, $\beta$ is a coupling constant and $D_{\rho}$ is the Weyl co-covariant derivative \cite{foot:Weyl}. Further \qo{matter fields}, as the Yang-Mills fields do not need to be explicitly expressed in Eq.~(\ref{eq:L}) since their coordinates can be accounted for, e.g., in a compactified form via a standard  Kaluza-Klein technique, by a multi-dimensional $D>4$ Weyl curvature~\cite{SalamStrathdee1982}. The generic potential $V(T,\phi,\mu)$, understood as an effective \qo{cosmological constant} function of the temperature $T$, accounts for the self-interaction of the scalar field $\phi$~\cite{PenroseHawking1996}. As we shall see immediately, the stringent conditions implied by the Weyl symmetry on each co-covariant addendum $X$ appearing in the expression of $L$ in Eq.~(\ref{eq:L}), i.e. $W(L)=W(X)=0$, impose a well defined exponential expression on the function $V(T,\phi,\mu)$.

In addition to the above considerations, we must further impose in the present context the general condition that the action: $I_\phi=\int{dx^4\sqrt{-g}L_\phi}$ is stationary respect to variations in $\phi$, where: $L_\phi \equiv \left[\frac{1}{2}g^{\rho\sigma}\partial_\rho\phi\partial_\sigma\phi-V(T,\phi)\right]$  expresses the effect of the inflation field in the $D=4$ space-time (Weinberg, 2008). This condition is expressed by:  
\begin{equation}\label{eq:Beltrami}
\nabla_B\phi = -\frac{\partial V(\phi,T)}{\partial\phi}
\end{equation}
being $\nabla_B$ the Laplace-Beltrami differential operator (Refs. 5 and 8).\\  

Let us now make the simplifying assumption that the inflationary system is "trapped" in local minimum of the potential, a significant dynamical condition largely considered in the literature~\cite{Dodelson2003}~\cite{foot:Lagrange} :$V'\equiv\frac{\partial V(\phi,T)}{\partial\phi}=0$. The variation of $L$ respect to the relevant fields  leads to the dynamical equations of the Weyl conformal theory. This one is selected to be integrable, so that $\phi_{\rho\sigma}=0$. In particular, the variation respect $g_{\rho\sigma}$ leads for $\alpha=\frac{1}{7}\simeq 0.1428$ to the following Einstein equation~\cite{Lord1979}:
\begin{equation}\label{eq:Einstein}
  R_{\rho\sigma} -\frac{1}{2}g_{\rho\sigma}R_g =\left[\partial_\rho\phi\partial_\sigma\phi -\frac{1}{2}g_{\rho\sigma}\partial_\eta\phi\partial^\eta\phi+g_{\rho\sigma}V(T,\phi)\right]
\end{equation}
being $R_{\rho\sigma}$ and $R_g$ the Ricci tensor and the curvature scalar of the Riemann geometry. The expression within square brackets at the  r.h.s. of Eq.~(\ref{eq:Einstein}) may be cast in the standard form:  $K^{2}T^*_{\rho\sigma}\equiv{K^{2}\left[\partial_\rho\phi\partial_\sigma\phi - g_{\rho\sigma}L_\phi\right]}$  with: $K^{2}=\frac{8\pi G}{c^4}$, by the rescaling: $\phi{\rightarrow{K\phi}}$ and:  $ V(\phi){\rightarrow{K^{2}V(K\phi)}}$ [i.e.:$\phi_{new}=K^{-1}\phi_ {old}].$ 
The Eq.~(\ref{eq:Einstein}) is a somewhat intriguing, suggestive, result  since it reproduces exactly the basic Einstein equation of the inflation theory reported in the standard texts on cosmology~\cite{Dodelson2003,Weinberg1972,PenroseHawking1996}. Note that our expression of $T^*_{\rho\sigma}$  is obtained, within the quoted simplifying assumption, by the formal application of the standard Euler-Lagrange variational procedures to the expression in Eq.~(\ref{eq:L}) of the $D=4$ Weyl's scalar curvature which is absent in Riemann's geometry~\cite{foot:Weyl}. Indeed, in modern texts an equation similar to Eq.~(\ref{eq:Einstein}) is attained via the introduction of an exotic\qo{quintessence} object wich is assumed to represent artificially a modified matter model~\cite{Amendola2010}. As said, by our novel interpretation of the inflaton field, the present work suggests the geometrical nature of the \qo{quintessence}. We shall see later in the paper that this leads to several consequences of  dynamical relevance, as for instance a straightforward, dynamically driven interchange between the weylan to riemannian symmetry conditions~\cite{HochbergPlunien1991}. Note that all the variational calculations e.g. leading to Eq (4). are carried out by considering the explicit spacetime dependence of the field: $\mu(x)$. Only after the completion of the variational process we have exploited the conformal, i.e. gauge invariance of the theory by choosing $\mu$ to be a \textit{constant} field. \\
As anticipated, the Weyl symmetry imposes a severe restriction to the explicit form of the massive \qo{inflaton potential}.  In virtue of Eq.~(\ref{eq:gauge}), since for any physical quantity: $X{\rightarrow{e^{\lambda(x)W(X)}X}}$, and because here: $W(V)=-2, W(\mu^{2})=-2, W(\sqrt{-2})=+4$, its general expression to be inserted in Eq.~(\ref{eq:L}) may be cast in the co-covariant form as a superposition of exponentials:
\begin{equation}\label{eq:V}
  V(T,K\phi,\mu)=\mu^2\sum_{n\ge1}C_n(T)(e^{-K\phi}/\mu)^{2n} 
\end{equation}
This generalization is interesting since $V$ may show an unlimited number of maxima or minima for $\phi$ ranging from zero to infinity~\cite{Amendola2010}. We notice nevertheless that in general $T$, and hence any $C_n(T)$, is not constant within the evolution of the inflaton $\phi$, e.g. during the rolling down of the Universe system towards levels of lower free energy in an expansion phase or, on the contrary, in a phase of re-heating.  In spite of these complications, we may easily recognize that Eq.~(\ref{eq:V}) is able to account reasonably for the so far inscrutable paths of the Universe evolution.\\

Let us consider,  for instance,  the first two terms of of the sum in Eq.~(\ref{eq:V}) and assume for simplicity that $C_1(T)$ and $C_2(T)$ are constant. We may then inquire about a path showing (for $C_2(T)<0$) a maximum of $V(T,K\phi)$ at $\phi=0$, i.e. $\partial V(T,0)/\partial\phi=V'(T,0)=0$. We may check that the plateau of this maximum is followed by an exponential rolling down of the system toward a \qo{true vacuum} at very large $\phi$.  A slightly more interesting example is offered by an initial fast decay of the system at $K\phi=0$. Assume for exaple that $\mu^{-2}C_2(T)=-C_1(T)$ i.e. $V(T,0)=0$. The system decays in a deep minimum followed by a re-heating phase towards $V(T,K\phi)=0$, attained at large values of $\phi$. In practice, as said, the lack of constancy of $T$ renders the overall evolution very complicated.\\

Since the time-time and space-space components for a homogeneous field $\phi$ are, respectively, $T^{*0}_0=-\rho$, $T^{*i}_i=P$, we may write
\begin{eqnarray}\label{eq:rhoP}
          \mbox{Energy density: }\rho_\phi &=&\frac{1}{2}\dot\phi^2+V(T,\phi) \nonumber\\
          \mbox{Pressure: }P_\phi &=& \frac{1}{2}\dot\phi^2-V(T,\phi)
 \end{eqnarray}
where, for any scalar $z$ we set $\dot z = z_{,0}=\partial_tz$ with $t$ cosmic time~\cite{Dodelson2003}. In the Robertson-Walker (RW) metric we may obtain the equation of motion for the inflaton via the Euler-Lagrange equation $\ddot\phi+3H\dot\phi+V'(T,\phi)=0$. Note that in most cases Eqs.(\ref{eq:rhoP}) in the RW metric can be solved in closed form in virtue of the explicit form of $V$ given by Eq.~(\ref{eq:V}). A field configuration with negative pressure implies a larger potential energy than kinetic $\dot\phi^2/2\ll V\sim$~constant, or $\ddot\phi\ll V'$. This condition may corresponds to a very slow roll of the inflaton $\phi$ on the plateau  of  the potential, or more interesting, when $\phi$ is trapped, i.e. with little kinetic energy, in a \textit{false vacuum}, i.e. in a \qo{local}, but not \qo{global} minimum of the potential. The \qo{trapping} condition is interesting because $\phi=$~const. implies a constant energy density, which is all potential~\cite{Dodelson2003}. However, a constant $\rho_\phi$ in an expanding universe, where the densities of both matter and radiation fall off very rapidly, implies that the universe becomes quickly dominated by the \textit{vacuum energy}. Moreover, it is worth noting in the context of our present theory that in a homogeneous and isotropic universe a zeroth-order trapped inflaton, i.e. $\phi^{(0)}$ with $\phi^{(0)}_{,0}=0$ and $\phi^{(0)}_{,i}=0$ $(i = 1,2,3)$, implies a zero Weyl curvature in the original Lagrangian, i.e. $R_W=0$. In this condition the highly symmetrical conformal Weyl gravitation theory reduces to the lower-symmetry Einstein's GR theory~\cite{HochbergPlunien1991}. This change of the dynamical mechanism with the roll down from an original very high temperature state of the representation point of the Universe along the $V(T,\phi)$ curve appears to be in agreement with the above statements about the possibility of a temperature driven symmetry-change which leads to the Riemannian structure we perceive today.\\
The general relevance of conformal symmetry and of the symmetry-breaking of the vacuum in the today quantum gravity endeavor is well emphasized by the following excerpt from a recent paper by Gerhard t'Hooft~\cite{Hooft2015}:\\
\qo{\textit{Why then is physics so difficult? Well, we still do not know what happens at higher energies even if we do understand the laws at low energies. Or more to the point: small time and distance scales seem not to be related to large time and distance scales. Now, we argue, this because we fail to understand symmetry of the scale transformations. This symmetry, of which the local form will be local conformal symmetry, if exact, should fulfill our needs. Since the world appears not to be scale invariant, this symmetry, if it exists must be spontaneously broken: This means that the symmetry must be associated with further field transformations leaving the vacuum not invariant. It is the implementation of the symmetry that we should attempt to construct from the evidence we have. In conclusion, there must be a component in space-time symmetry group (the Poincar\'{e} group) that both Lorentz and Einstein dismissed\dots}}.\\

The Title of the paper from which the last excerpt was taken is self-explanatory: \qo{\textit{Local conformal symmetry: the missing symmetry component for space and time}}. We believe that the content of the present work fits rather neatly some relevant aspects of the  program envisaged by t'Hooft.\\

In addition to the above considerations it is appealing to consider that the geometrical mechanism proposed in the present work represents a relevant unifying scenario by which the same \qo{inflaton} scalar field appears to play an essential role in determining the evolution of the Universe \qo{at large} as well as, at the microscopic level and via the dynamics of the Weyl scalar curvature $R_W$,  of the everyday quantum phenomenology~\cite{foot:QCD}.This appears to be a glimpse on quantum gravity~\cite{De Martini2016}.\\

\begin{thebibliography}{24}
\bibitem[{foo({\natexlab{b}})}]{foot:Weylerror}
\bibinfo{note}{The correct application of co-covariance was overlooked by Weyl
himself after discovering his erroneous interpretation of the electromagnetic field
as expressed by the quantity $\phi_{\rho\sigma}$.}
\end{thebibliography}

\end{document}